\documentclass[aps,prb,twocolumn,superscriptaddress]{revtex4}
\usepackage{graphicx}
\usepackage{times}
\usepackage{amsmath}
\usepackage{textcomp}
\usepackage{epstopdf}
\usepackage{float}

\begin{document}
\title{ Nonlocal Andreev reflection, fractional charge and current-phase relation in topological bilayer exciton condensate junctions}
\author{M. Veldhorst}
\affiliation{ARC Centre of Excellence for Quantum Computation and Communication Technology, School of Electrical Engineering \& Telecommunications, The University of New South  Wales, Sydney 2052, Australia}
\author{M. Hoek}
\affiliation{Faculty of Science and Technology and MESA+ Institute for Nanotechnology, University of Twente, 7500 AE Enschede, The Netherlands}
\author{M. Snelder}
\affiliation{Faculty of Science and Technology and MESA+ Institute for Nanotechnology, University of Twente, 7500 AE Enschede, The Netherlands}
\author{H. Hilgenkamp}
\altaffiliation[Also at ]{Leiden Institute of Physics, Leiden University, P.O. Box 9506, 2300 RA Leiden, The Netherlands}
\affiliation{Faculty of Science and Technology and MESA+ Institute for Nanotechnology, University of Twente, 7500 AE Enschede, The Netherlands}
\author{A.A. Golubov}
\altaffiliation[Also at ]{Moscow Institute of Physics an Technology, Dolgoprudnyi, Moscow district, Russia}
\affiliation{Faculty of Science and Technology and MESA+ Institute for Nanotechnology, University of Twente, 7500 AE Enschede, The Netherlands}
\author{A. Brinkman}
\affiliation{Faculty of Science and Technology and MESA+ Institute for Nanotechnology, University of Twente, 7500 AE Enschede, The Netherlands}
\date{\today}
\begin{abstract}
We study Andreev reflection and Josephson currents in topological bilayer exciton condensates (TEC). These systems can create 100\% spin entangled nonlocal currents with high amplitudes due to perfect nonlocal Andreev reflection. This Andreev reflection process can be gate tuned from a regime of purely retro reflection to purely specular reflection. We have studied the bound states in TEC-TI-TEC Josephson junctions and find a gapless dispersion for perpendicular incidence. The presence of a sharp transition in the supercurrent-phase relationship when the system is in equilibrium is a signature of fractional charge, which can be further revealed in ac measurements faster than relaxation processes via Landau-Zener processes.
\end{abstract}
\maketitle
Fermionic condensates have been intensively studied since its first experimental discovery in superconducting mercury more than a century ago \cite{Onnes}. These condensates bear spectacular effects, such as macroscopic phase coherence and magnetic flux quantization. A lot of research has been devoted to the superconducting class of condensates and with the appearance of nanotechnology, which has enabled the study of interfaces at the nanoscale, many new exciting experiments have been proposed. Strong similarities between superconducting and exciton condensates, which arises from the Coulomb interaction between electron and holes \cite{Shevchenko1976, Lozovik1997}, have been recognised early on, but with the recent advances in bilayer exciton condensates it becomes particularly interesting to study superconducting effects in exciton condensates.

Exciton condensation has been realized in closely spaced quantum Hall bilayers \cite{Eisenstein2004}, enabling studies towards Andreev reflection and Josephson effects \cite{Finck2011, Nandi2012, Lagoudakis2010}. Topological exciton condensation has been predicted in three dimensional topological insulators, potentially surviving up to room temperature \cite{Seradjeh2009}. The topological exciton condensate (TEC) arises in this case from the pairing of carriers mediated by the Coulomb interaction in closely spaced top and bottom surfaces of a topological insulator (TI). The strong experimental progress in tuning the Fermi energy inside the bulk bandgap bears promise for the realization of these systems \cite{Jun2012, Ren2011}. Motivated by the similarities between superconducting and excitonic condensates, we discuss here key superconducting phenomena and show how these effects manifest themselves in the topological exciton condensate.

\begin{figure} [!ht]
	\centering 
		\includegraphics[width=0.5\textwidth]{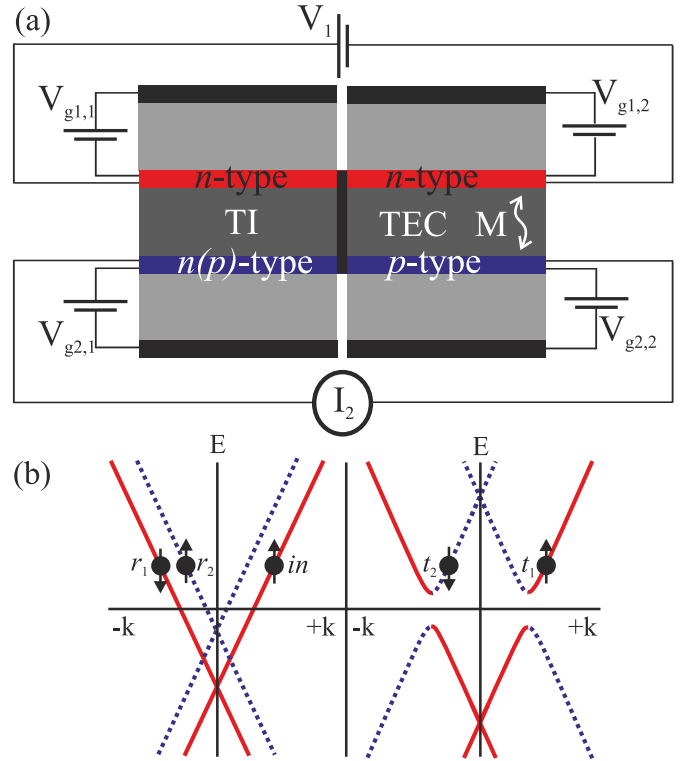}
		\caption{ (a) TI -TEC  heterostructure with individual gates to tune between $n$ and $p$ type surface states. The right side forms an exciton condensate due to the Coulomb interaction between $n$- and $p$-layers. Applying a voltage $V_1$ over the top surface states, creates a nonlocal current $I_2$ through the bottom surface states. (b) Allowed transport processes in the device (the TI is considered here to be of $nn$-type); the arrows denote the spin normal to the interface, $\hat{\textbf{e}}_x$, and solid red (dotted blue) indicates the top (bottom) surface, respectively. An incoming top surface electron ($in$) can be reflected ($r_1$) and Andreev reflected ($r_2$) as an electron. Transmission from the TI to the TEC occurs as quasiparticles with electron-like mass ($t_1$) and hole-like mass ($t_2$). Elastic co-tunneling has a vanishing probability due to the large intrinsic TI bulk bandgap.}  
		\label{fig:TEC1}
\end{figure}
Electron reflection at an interface is an intensively studied quantum phenomenon. One of the most famous examples is the electron-hole Andreev reflection occurring at the superconductor interface. Usually, normal electron scattering occurs via specular reflection, due to translational invariance along the interface, whereas Andreev reflection is of retro-type, due to the sign change in the group velocity. Recently, it has been predicted that in special cases these processes can be of the opposite type. Electron retro reflection is predicted at the interface of a superconductor with bilayer graphene \cite{Ang2012}. The Andreev reflection process becomes partially specular when the material contacting the superconductor has a gapless dispersion, such as graphene and topological insulators, and is tuned to the regime where the incoming electron has energy above the Dirac point and the retro-reflected hole has energy below the Dirac point \cite{Beenakker2006}. These novel processes attracted great attention, but are yet to be observed. Here, we show that Andreev reflection in exciton condensates can be tuned from completely retro-reflection to completely specular reflection purely by electrical gating.

Superconductors are a natural source of entanglement; the Cooper pair charge carriers in $s$-wave superconductors are in a singlet state. Most proposals using superconductors to create nonlocal entangled electrons are based on splitting the Cooper pairs via crossed Andreev reflection \cite{Byers1995, Deutscher2000, Falci2001, Hermann2010, Hofstetter2009, Veldhorst2010}. However, the current is often only for a small part entangled due to the competing processes of normal Andreev reflection and elastic co-tunneling \cite{Falci2001}. Proposals to optimize crossed Andreev reflection have focused on the electrodes contacting the superconductor. The fraction of entangled particles can be strongly increased by using ferromagnetic electrodes in an antiparallel magnetization \cite{Deutscher2000}, and could even reach 100$\%$ in a $p$-type semiconductor - superconductor - $n$-type semiconductor junction \cite{Veldhorst2010}. Still, these proposals rely on very specific configurations and are always limited by the critical temperature of the superconductor. Here, we show that Andreev reflection on bilayer exciton condensates is naturally nonlocal. The Andreev reflection amplitudes are high in the presence of spin-momentum locking, which is the case in topological condensates. These results bear promise for the realization of ideal entanglement sources.

A superconducting Josephson junction is predicted to host zero energy bound states if the interlayer is made out of topological insulators \cite{Fu2008}. These modes are Majorana modes as the superconductor provides the right particle-hole symmetry and the topological insulator makes the quasiparticles spinless. The search for Majorana zero energy modes is of practical relevance as these particles might serve as decoherence immune qubits with non-Abelian statistics \cite{Moore1991} in topological quantum computation \cite{Kitaev2003}. These Josephson junctions show an exotic supercurrent-phase relationship and can have a doubled periodicity. Here, we show that Josephson effects arise by coupling two TECs. The transparency of the Josephson junction is angle dependent and the bound states are gapless for perpendicular incidence. We find that these bound states have no parity protection due to degeneracy in layers, contrary to topological superconductors. The zero energy bound states are not Majorana modes, due to degeneracy, similar to the valley degeneracy in graphene. We have calculated the supercurrent and find a sharp transition when the system is in equilibrium in the current-phase relationship around $\phi=\pi$ for perpendicular incidence. This transition is a signature of currents quantized in fractional charge. AC measurements can reveal fractional charge and current phase relationships due to Landau-Zener processes.

Coulomb interaction can induce exciton condensation when the two surfaces of a topological insulator material, which is insulating in the bulk with a finite bandgap \cite{Zhang2006, Fu2007}, are sufficiently close\cite{Seradjeh2009}, shown in Fig. \ref{fig:TEC1}. All layers are assumed to be individually tunable by means of electrical gates. The electrical gates attached to the exciton condensate are used to tune the top (bottom) surface of the topological insulator to be of $n(p)$-type, resulting in an attractive Coulomb interaction $\hat{M}$, that drives the system to exciton condensation. Strong coupling is expected which may survive up to room temperature \cite{Seradjeh2009, Wang2012}. The linear energy dispersion of the topological surface states cause a near perfect nesting between the electron-like states above the Dirac point and the hole-like states below the Dirac point \cite{Seradjeh2009}. This situation is similar to the prediction of exciton condensation in graphene \cite{Lozovik2008, Min2008}, except that graphene has an additional pseudospin. The two dimensional nature of the surface states reduces screening and maximizes the Coulomb interaction. 

\begin{figure}
	\centering 
		\includegraphics[width=0.5\textwidth]{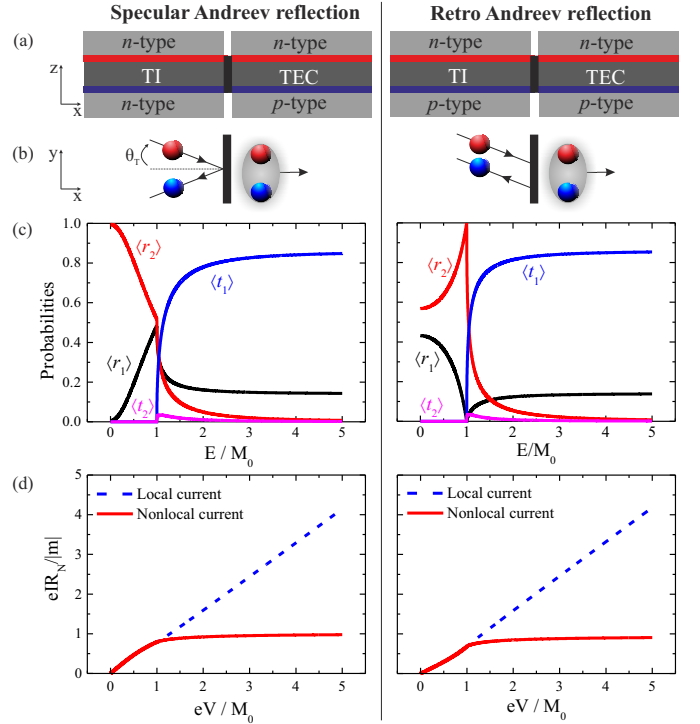}
		\caption{ Exciton Andreev reflection. On the left we show the limit of specular Andreev reflection and on the right the limit of retro Andreev reflection. (a) Electrode configuration to obtain the specific configurations. The TI-leads are of similar (opposite) type for specular (retro) reflection. (b) Conservation of parallel momentum, together with a group velocity pointing away from the interface results in specular reflection as the top and bottom leads are of the same charge type and retro reflection when the leads have opposite charge type. (c) Angle averaged tunneling coefficients and (d) $IV$ characteristics. Electrons with energy $|E|<M_0$ can only enter the exciton condensate by the exciton analogue of Andreev reflection. For energies $|E|>M_0$, also quasiparticle current appears. The blue dashed line is the current through the same interface where the voltage is applied; the red solid line is the resulting nonlocal current at the other interface. At $|eV|<M_0$ the current is perfectly entangled in both scenarios. }  
		\label{fig:TEC2}
\end{figure}

The surface states of a topological insulator can be described by
\begin{equation}
\hat{H}+\mu_{T(B)}\hat{I}=+(-)v_{T(B)} \boldsymbol\sigma \cdot \hat{\textbf{p}}.
\label{eq:TI}
\end{equation}
Here, the momentum $\hat{\textbf{p}}=-i\hbar\nabla$ of the topological insulator is coupled to the spin, and $\boldsymbol\sigma=(\sigma_x,\sigma_y)$ are the Pauli spin matrices. The Fermi velocity $v_{T(B)}$ represents the Dirac velocity in the top ($T$) and bottom ($B$) layer, and the (+)(-) is due to the different chiralities residing at the two sides. The Fermi energy $\mu_{T(B)}=\mu_0+E_{pT(B)}$ is given by the intrinsic chemical potential $\mu_0$ and the potential energy $E_p$, tuned by the electrical gates. The Hamiltonian after mean field approximation is, in the basis $(\hat{c}_{Tk\uparrow},\hat{c}_{Tk\downarrow},\hat{c}_{Bk\uparrow},\hat{c}_{Bk\downarrow})$, given by  \cite{Seradjeh2009}
\begin{equation}
\hat{H}+\mu_0\hat{I}=
\left(
\begin{array}{cc}
v_T \boldsymbol\sigma \cdot \hat{\textbf{p}}-E_{pT} & \hat{M} \\
\hat{M}^{*} & -v_B \boldsymbol\sigma \cdot \hat{\textbf{p}}-E_{pB} \end{array}
 \right).
 \label{eq:TEC}
\end{equation}
The Coulomb interaction is not directly spin-selective, and the form of $\hat{M}$ will depend on the actual system. For a diagonal interaction in spin-space, the exciton orderparameter $\hat{M}=M \left\langle \hat{c}_{Tks}^{\dagger}\hat{c}_{Bks}\right\rangle + h.c.$, where we assume $M=M_0 e^{i\phi}$ and $s$ denotes the spin. The TEC condensate phase $e^{i\phi}$ will be important when coupling different exciton condensates.
The eigenvalues corresponding to this system are given by
\begin{equation}
\begin{array}{ll}
E_{k\alpha \eta}  = &  -\mu+\frac{1}{2}(E_{p,T}+E_{p,B}) \\
& + \alpha \sqrt{\left[v|k|+\frac{1}{2}\eta(E_{p,B}-E_{p,T})\right]^2+M_0^2}.
\end{array}
\end{equation}
Here, $\alpha,\eta=\pm1$. We will focus on the regime $\frac{1}{2}(E_{p,2}-E_{p,1})-\mu\rightarrow0$, where the condensation energy is maximized. We attach the TEC to normal topological insulator electrodes, described by Eq. \ref{eq:TI}. This setup is shown in Fig. \ref{fig:TEC1}. All layers can be tuned individually by means of electrical gates. As Seradjeh $et$ $al.$ \cite{Seradjeh2009} pointed out, the exciton condensation energy $M_0$ will vanish for a small Fermi energy difference $E_{p,T}-E_{p,B}$ and a large mean Fermi energy potential $\mu+\frac{1}{2}(E_{p,T}+E_{p,B})$. Consequently, we can neglect the Coulomb interaction in the $nn$ configuration. We also neglect the Coulomb interaction for the TI in the $pn$ configuration when using low carrier densities and large density mismatches. In the rigid boundary approximation this results in the following ansatz
\begin{eqnarray}
\nonumber
\hspace{-5mm}&& \Psi_{C} \hspace{-0.5mm} = \hspace{-0.5mm} t_1 \hspace{-1mm} \begin{pmatrix}  u e^{i\phi} \\ u e^{i\phi+i\theta_{CT}}\\ v e^{-i\phi}\\ \hspace{-1mm} -v e^{-i\phi-i\theta_{CT}} \hspace{-1mm} \end{pmatrix} \hspace{-0.5mm} e^{i \textbf{k}_{CT} \textbf{r}} \hspace{-0.5mm} + \hspace{-0.5mm} t_2 \hspace{-1mm} \begin{pmatrix} v e^{i\phi} \\ -v e^{i\phi-i\theta_{CB}} \\u e^{-i\phi }\\ \hspace{-1mm} u e^{-i\phi+i\theta_{CB} \hspace{-1mm} } \end{pmatrix} \hspace{-1mm} e^{-i \textbf{k}_{CB} \textbf{r}} \\
\nonumber
\hspace{-5mm}&& \Psi_T  \hspace{-0.5mm}=\hspace{-1mm} \begin{pmatrix} 1 \\ \hspace{-1mm} e^{i \theta_T} \hspace{-1mm} \\ 0 \\0  \end{pmatrix}  \hspace{-1mm} e^{i \textbf{k}_{T} \textbf{r}}+ \hspace{-0.5mm} r_1 \hspace{-1mm} \begin{pmatrix} 1 \\\hspace{-1mm} -e^{-i \theta_T} \hspace{-1mm} \\ 0 \\0  \end{pmatrix} \hspace{-1mm} e^{-i \textbf{k}_{T} \textbf{r}} \\
\hspace{-5mm}&&  \Psi_B \hspace{-0.5mm}= \hspace{-0.5mm} r_2 \begin{pmatrix} 0 \\ 0 \\ 1 \\\ \hspace{-1mm} -e^{i \beta \theta_B} \hspace{-1mm}\end{pmatrix} \hspace{-1mm} e^{-i \beta \textbf{k}_{B} \textbf{r}} 
\label{eq:RT} 
\end{eqnarray}
Here, $\hat{\Psi}_{T(B)}$ is the wave function in the top (bottom) surface, $\theta$ the trajectory angle (see Fig. \ref{fig:TEC2}), $k$ the momentum, $r,t$ the probabillity coefficients, and $\hat{\Psi}_C$ refers to the bilayer exciton condensate. For general parameters, $u$ and $v$ are found through the Hamiltonian, Eq. \ref{eq:TEC}, together with demanding $\sum_{i}{|\hat{\Psi}_i|^2}=1$. For equal Dirac velocities and carrier densities in the exciton layer, the coherence factors $u$ and $v$ are given by $u=\sqrt{\frac{1}{2}+\frac{1}{2}\frac{\sqrt{E^2-M_0^2}}{E}}$ and $v=\sqrt{1-u^2}$.

A particle in the TI layer impinging on the TEC can have several elastic scatter trajectories, with the angle $\theta_j$ of a scatter  trajectory $j$, defined with respect to the interface, related to the incoming angle $\theta_{in}$ via Snell's law $\sin(\theta_j)=r_k\sin(\theta_{in})$ with $r_k=\frac{k_{in}}{k_j}$ because of the conservation of momentum parallel to the interface. A particle can backscatter in the TI, with probability $r_1$, while changing its spin accordingly. It can also undergo Andreev reflection ($r_2$) by scattering to the other TI layer. It will have opposite (same) perpendicular momentum, when the TI electrodes are of $nn$-type or $pn$-type, respectively. Therefore, Andreev reflection is specular (retro) when both electrodes are of similar (opposite) type, in contrast to normal metal superconductor contacts. Parallel momentum is conserved, resulting in $\beta =+(-)1$, see Eq. \ref{eq:RT}, for (specular) retro reflection due to the different chirality between the $p$ and $n$ configuration. Therefore, the spin of the Andreev reflected electron is dependent on whether the reflection is retro or specular. In the exciton system, tuning from specular to retro can be achieved by tuning the gate voltages, while in normal metal - superconductor contacts specular Andreev reflection is predicted only for very specific cases \cite{Beenakker2006}. 
A particle in the TI layer can also scatter as a quasiparticle into the TEC. This transmission from the TI into the TEC is possible via scattering into the electron-branch ($t_{1}$) or hole-branch ($t_{2}$), but is absent for excitation energies smaller than the exciton gap $M_0$. Direct tunneling of particles between the top and bottom layers of both the TI and TEC is not taken explicitly in the model, as direct tunneling decays very rapidly with increasing layer separation distance $d$. The bulk bandgap $\Delta_{TI}>$100 meV and the decay scales as $e^{- k d}$, with $k\propto \Delta_{TI}$.  A regime where the dominant process is nonlocal Andreev reflection is therefore easily obtained, whereas this optimization in superconducting systems is a major hurdle. 

In obtaining the scatter possibilities, we assume $\hat{\Psi}$ to be continuous across the interface. We integrate the probability distribution over angles $\theta_{in}$ from 0 to $\pi$, considering a step-like interface along the direction  $\hat{\textbf{e}}_x$ normal to the interface. Figure \ref{fig:TEC2} shows the angle averaged scatter probabilities for $\kappa=\frac{\mu_{TI}}{\mu_{TEC}}=0.1$ with the electrodes in the $nn$ and $pn$ configuration, which is representative of the general result for a large chemical potential mismatch, since $\theta_{t1,t2} \rightarrow 0$ for $\kappa \rightarrow 0$. When the TI electrodes are in the $nn$ configuration, Andreev reflection is specular and peaks at zero energy, similar to what is predicted for graphene \cite{Beenakker2006}. Backscattering is forbidden on the edge of a 2D topological insulator, but scattering at other angles apart from $\pi$ is possible on the 2D surface of a 3D topological insulator, leaving a nonzero electron reflection. Still, the obtained Andreev reflection $r_2$ is significant, and will increase for smaller chemical potential mismatches. Effectively, the interface has a high transparency for all mismatches. 

The current density in the electrodes in the $\hat{\textbf{e}}_x$-direction, perpendicular to the interface, is obtained from $J_{x,T(B)}=\frac{1}{A} \sum_{\textbf{k}} \textbf{J}_{q,T(B)}(\textbf{k})\hat{\textbf{e}}_xf_{T(B)}(\textbf{k})$. Here, $A$ is the effective width, and the nonequilibrium distribution $f_{T(B)}=f_0(E-eV_T(V_B))-f_0(E-eV_{ex,T(B)})$, with $f_0(E)$ the Fermi distribution function. Only trajectories below the critical angle $\theta_c(E)=\arcsin{r_k^{-1}}$ contribute to the current. The charge current is defined by $\textbf{J}_{q,T(B)}=ev_{T(B)}[\Re(\hat{\Psi}_\uparrow\hat{\Psi}_\downarrow^*)\hat{\textbf{e}}_x+\Im(\hat{\Psi}_\uparrow\hat{\Psi}_\downarrow^*) \hat{\textbf{e}}_y]$. Bias voltages in the range $|eV|<M_0$ result in vanishing quasiparticle current in the TEC, and direct tunneling is negligible in this system. Therefore, in this regime, perfect entangled currents flow through both surfaces in opposite directions, see the lower panels of Fig. \ref{fig:TEC2}. The currents are entangled in energy, momentum and spin and form a promising source for solid state Bell experiments, quantum computation and quantum teleportation.  The spin-momentum locking provides furthermore an additional path to probe the entanglement \cite{Sato2010}.
\begin{figure}
	\centering 
		\includegraphics[width=0.5\textwidth]{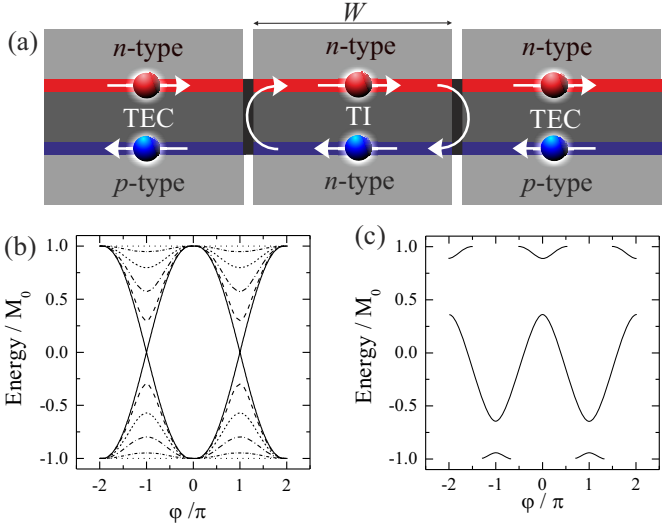}
		\caption{ (a) Bilayer exciton condensate analogy of the Josephson junction. The arrows indicate the direction of the group velocity. The group velocity is in the same (opposite) direction as the momentum in the $n(p)$-type branches . A possible exciton bound state is shown. (b) The boundstate for perpendicular incidence (solid line) is 4$\pi$ periodic. Nonzero incidence angle results in the opening of a gap at finite length ($\frac{W}{\xi}=0.1 $ here) and momentum mismatch ($r_k=0.1$), as shown for $\theta_T=0.1$, 0.2, 0.3, 0.4 and $0.5 \pi$ in dashed lines. (c) When the top and bottom TI layers have unequal Fermi densities, the gap shifts from zero energy, here $E_{p,B}=2.6E_{p,T}$, $W=0.1$, $\theta_T=0.2\pi$ and $r_k=0.1$.}  
		\label{fig:TEC3}
\end{figure}
\begin{figure}
	\centering 
		\includegraphics[width=0.5\textwidth]{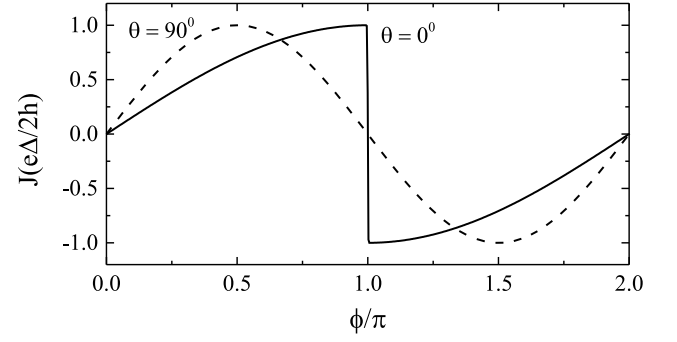}
		\caption{ Exciton Josephson supercurrent phase relationship in equilibrium. The limit of parallel incidence (normalized) results in a 2$\pi$ periodicity due to the presence of a gap in the bound states (Fig. \ref{fig:TEC3}). However, a gapless dispersion for perpendicular incidence moves the maximum supercurrent to $\phi=\pi$ and is the on-set of a doubled periodicity and fractional charge. Relaxation causes a sharp transition around $\phi=\pi$, where the current switches between the two branches.}  
		\label{fig:TEC4}
\end{figure}

We consider two TECs connected by a topological insulator with width $W$ as analogue to a superconductor Josephson junction, see Fig. \ref{fig:TEC3}a. As for a single interface, see Fig. \ref{fig:TEC1}, a current can be applied and measured across one layer (e.g. the top layer), or measured nonlocally via the other layer. In the topological insulator electrodes there are no pairing interactions, but the Andreev reflected particle remains coherent with the incoming particle over a length $\xi=\frac{\hbar v_T v_B}{C_1+C_2}$. The factor $C_1=(v_T+v_B)E$ is the consequence of condensation of particles with energy $E$ above and below the chemical potential. The additional factor $C_2=(v_T-v_B)\mu_0+(v_B+v_T \frac{E_{pB}}{E_{pT}})E_{pT}$ is due to possible differences in Fermi velocity and energy between the two layers. In the case when both layers have equal electron densities and velocities, the characteristic phase coherence length is maximized and can be written as $\xi=\hbar v_D/M_0$ (similar to the superconducting coherence length by substituting $M_0$ with the superconducting gap $\Delta$). When the width of the TI interlayer $W \approx \xi$, an exciton supercurrent can flow between the two TECs. 

To find the exciton bound states, we use the ansatz: $\hat{\Psi}=a \hat{\Psi}_T^+ +b \hat{\Psi}_B^+ +c \hat{\Psi}_T^-+d \hat{\Psi}_B^-$, where superscript $\pm$ denotes forward and backward traveling waves. The bound states for this system are solved by assuming $\hat{\Psi}$ to be continuous across both the left ($L$) and right ($R$) interfaces. There are several ways to calculate the bound states; we follow the approach of Kulik \cite{Kulik1970}. The modes are calculated for energies $E<M_0$. The system is solved by connecting the left and right moving currents
\begin{eqnarray}
\hspace{-5mm}&&c \hat{\Psi}_T^-(W,E)           =   r_1^{T,R}(E)a \hat{\Psi}_T^+(W,E)+r_2^{B,R}(E)d \hat{\Psi}_B^-(W,E)  \nonumber \\ 
\hspace{-5mm}&&b \hat{\Psi}_B^+(W,E)  =   r_2^{T,R}(E)a \hat{\Psi}_T^+(W,E)+r_1^{B,R}(E)d \hat{\Psi}_B^-(W,E)  \nonumber \\ 
\hspace{-5mm}&&a \hat{\Psi}_T^+(0,E)           =  r_1^{T,L}(E)c \hat{\Psi}_T^-(0,E)+r_2^{B,L}(E)b \hat{\Psi}_B^+(0,E) \nonumber \\
\hspace{-5mm}&&d \hat{\Psi}_B^-(0,E)  =  r_2^{T,L}(E)c \hat{\Psi}_T^-(0,E)+r_1^{B,L}(E)b \hat{\Psi}_B^+(0,E). 
\label{system:currentflow}
\end{eqnarray}
The coefficients $r$ are determined by considering scattering at a single interface using the ansatz, Eq. (4). Figure \ref{fig:TEC3}b displays the boundstates for different incident angle $\theta_T$. Scattering present at finite angles results in the opening of a gap. For equal electron densities and perpendicular incidence, the absence of backscattering results in a zero energy state for any finite length $W$. Unequal electron densities in the top and bottom layer removes the particle-hole symmetry and shifts the gap from zero energy, resulting in zero energy bound state for all angles, see Fig. \ref{fig:TEC3}c. The zero energy state appears at different $\phi$ for different incidence angle, but the current phase relationship is always 2$\pi$ periodic.

The Andreev bound states in topological superconducting systems are protected by parity, resulting in a 4$\pi$ periodic current phase relation for perpendicular incidence \cite{Fu2009}. This doubling of the period is the consequence of a switch from Cooper pairs to single electrons of the transferred charge \cite{Beenakker2012}. The supercurrent in these topological exciton systems is carried by single electrons in the top and bottom layers, such that a doubling in period would be a switch to transferring fractional charges across the individual layers, where the system has to be advanced with 4$\pi$ in $\phi$ to return to its original state (and an electron transfer across the interface). However, in the considered exciton junction, there is degeneracy in layer, and this lifts parity protection. This becomes evident when we consider the two bound states at perpendicular incidence: $\pm \epsilon(\phi)=\pm \cos(\phi/2)$. If we take the inner product of the corresponding two eigenstates $\Gamma_{\pm}$, we get the effective low energy Hamiltonian $H_0 = \epsilon(\phi)\left[\Gamma^\dagger_+\Gamma_{+}-\Gamma^\dagger_{-}\Gamma_{-}\right]$. In topological superconducting systems it is possible to get $\Gamma_+^\dagger=\Gamma_-$ due to particle-hole symmetry. However, here $\Gamma_+^\dagger\neq\Gamma_-$ as the eigen states correspond to different surfaces. Consequently, the dispersive bound states, $\pm \epsilon(\phi)$, correspond both to odd occupied states and there is therefore no parity protection. 

The resulting dc Josephson supercurrent will likely be in the equilibrium regime, as there is no parity protection that forbids any matrix element to couple to the two branches. We therefore calculate the supercurrent by taking the derivative with respect to the Free energy, $J_{T,B}=\pm \partial_{\phi} T \ln \sum_i e^{-E_i}/T$, where $i$ denotes the branch. For temperatures $T \ll E_i$, the supercurrent will be carried by the ground state only. In Fig. \ref{fig:TEC4} we show the supercurrent for perpendicular and for the limit of parallel incidence. The supercurrent for parallel incidence is normalized for clarity. As the angle of incidence decreases, the gap in the bound states decreases, and the maximum supercurrent shifts towards $\phi=\pi$, accompanied by a sudden transition where the supercurrent switches between the two branches. This transition is the result of absence of parity protection, but it is, together with the shift of maximum supercurrent towards $\phi=\pi$ the onset of a doubled current-phase relation (from 2$\pi$ to 4$\pi$) and the transfer of fractional charge across each layer ($e/2$).

Although there is no strong protection for relaxation between the two low energy bound states, the doubling in period can be studied further in an ac measurement, performed faster than the relaxation processes. Via Landau-Zener transitions, in a non-equilibrium measurement, the current can remain in the same branch as $\phi$ is advanced. The exciton ac Josephson effect follows from $\delta_t \phi=q^* V$, where $q=e$ on a single layer. In a microwave irradiation experiment, Shapiro steps form as function of the applied voltage and are quantized in $V=\frac{2h}{e}f_{RF}$. 

Recently, fluxoid quantization is predicted in bilayer exciton systems \cite{Rademaker2011}, quantized in $\Phi_0^*=\frac{h}{e}\gamma$, with $\gamma$ the diamagnetic susceptibility. The doubled current phase relationship would double the quantization resulting in $\Phi_0^*=2\frac{h}{e}\gamma$, which could be observed in SQUID devices \cite{Veldhorst2012PRB}. We note that even in the presence of relaxation, these altered current-phase relationships can be measured in SQUID devices \cite{Veldhorst2013}. 

In conclusion, we have studied the coupling between topological exciton condensates and topological insulators. A single interface opens the possibility to create 100\% spin entangled and spatially separated particles via nonlocal Andreev reflection, where the spin-momentum locking introduces new means to read out the entanglement. This novel Andreev reflection can be tuned from retro-reflection to a regime of specular reflection purely by electrical gating. Sandwiching a topological insulator between topological exciton condensates in a Josephson junction arrangement results in Josephson supercurrent with a gapless dispersion for perpendicular incidence. Degeneracy in layer lifts the parity protection, and the supercurrent has a strong transition around $\phi=\pi$ in equilibrium, and is the onset of fractional charges. In ac measurements faster than the relaxation, the current-phase relationship can attain a doubled periodicity (from 2$\pi$ to 4$\pi$ in the phase $\phi$), which gives rise to Shapiro steps with height $V=\frac{2h}{e}f_{RF}$, four times larger than in a standard superconducting Josephson junction where charge is carried by Cooper pairs with charge $2e$. Given the strong activity to realize topological insulators with an insulating bulk and the demonstration of few layer thin film topological insulators makes this proposal particularly timely. 

This work is supported by the Australian Research Council Centre of Excellence for Quantum Computation and Communication Technology (Project No. CE11E0096), the U.S. Army Research Ofﬁce (Grant No. W911NF-13-1-0024), the Netherlands Organization for Scientiﬁc Research (NWO), the Dutch Foundation for Fundamental Research on Matter (FOM) and the Russian Ministry of Education and Science.


\begin{thebibliography}{9} 
\bibitem{Onnes} H.K. Onnes, Commun. Phys. Lab. \textbf{12}, 120 (1911).
\bibitem{Shevchenko1976} S.I. Shevchenko, Sov. J. Low Temp. Phys. \textbf{2}, 251 (1976).
\bibitem{Lozovik1997} Y.E. Lozovik and A.V. Poushnov, Phys. Lett. A \textbf{228}, 399 (1997).
\bibitem{Eisenstein2004} J.P. Eisenstein, A.H. MacDonald, Nature \textbf{432}, 691 (2004).
\bibitem{Finck2011} A.D.K. Finck, J.P. Eisenstein, L.N. Pfeiffer, and K.W. West, Phys. Rev. Lett. \textbf{106}, 236807 (2011).
\bibitem{Nandi2012} D. Nandi, A.D.K. Finck, J.P. Eisenstein, L.N. Pfeiffer, and K.W. West, Nature \textbf{488}, 481-484 (2012).
\bibitem{Lagoudakis2010} K.G. Lagoudakis, B. Pietka, M. Wouters, R. Andre, and B. Deveaud-Pledran, Phys. Rev. Lett. \textbf{105}, 120403 (2010).
\bibitem{Seradjeh2009} B. Seradjeh, J.E. Moore, and M. Franz, Phys. Rev. Lett. \textbf{103}, 066402 (2009).
\bibitem{Ren2011} Z.R. Ren, A.A. Taskin, S. Sasaki, K. Segawa, Y. Ando, Phys. Rev. B \textbf{84}, 165311 (2011). 
\bibitem{Jun2012} X. Jun, A.C. Petersen, D. Qu, Y.S. Hor, R.J. Cava, and N.P. Ong, Physica E \textbf{44}, 917 (2012).
\bibitem{Byers1995} J.M. Byers and M.E. Flatt\'e, Phys. Rev. Lett. \textbf{74}, 306 (1995).
\bibitem{Deutscher2000} G. Deutscher and D. Feinberg, Appl. Phys. Lett. \textbf{76}, 487 (2000).
\bibitem{Falci2001} G. Falci, D. Feinberg, F.W.J. Hekking, Europhys. Lett. \textbf{54}, 255 (2001).
\bibitem{Hermann2010} L.G. Herrmann, F. Portier, P. Roche, A.L. Yeyati, T. Kontos, C. Strunk, Phys. Rev. Lett. \textbf{104}, 026801 (2010).
\bibitem{Hofstetter2009} L. Hofstetter, S. Csonka, J. Nygard, and C. Sch$\ddot{\textrm{o}}$nenberg, Nature \textbf{461}, 960 (2009).
\bibitem{Veldhorst2010} M. Veldhorst and A. Brinkman, Phys. Rev. Lett \textbf{105}, 107002 (2010).
\bibitem{Fu2008} L. Fu and C.L. Kane, Phys. Rev. Lett. \textbf{100}, 096407 (2008).
\bibitem{Moore1991} G. Moore and N. Read, Nucl. Phys. \textbf{B360}, 362 (1991).
\bibitem{Kitaev2003} A. Kitaev, Ann. Phys. (N.Y.) \textbf{303}, 2 (2003).
\bibitem{Zhang2006} B.A. Bernevig, T.L. Hughes, and S.C. Zhang, Science \textbf{314}, 1757 (2006).
\bibitem{Fu2007} L. Fu, C.L. Kane, and E.J. Mele, Phys. Rev. Lett. \textbf{98}, 106803 (2007). 
\bibitem{Wang2012} Z. Wang, N. Hao, Z.G. Fu, P. Zhang, New J. Phys. \textbf{14}, 063010 (2012).
\bibitem{Lozovik2008} Y.E. Lozovik and A.A. Sokolik, JETP Lett. \textbf{87}, 55 (2008).
\bibitem{Min2008} H. Min, R. Bistritzer, J.J. Su, and A.H. MacDonald, Phys. Rev. B \textbf{78}, 121401(R) (2008).
\bibitem{Read2000} N. Read and D. Green, Phys. Rev. B \textbf{61}, 10267 (2000).
\bibitem{Tanaka2009} Y. Tanaka, T. Yokoyama, and N. Nagaosa, Phys. Rev. Lett. \textbf{103}, 107002 (2009). 
\bibitem{Beenakker2006} C.W.J. Beenakker, Phys. Rev. Lett. \textbf{97}, 067007 (2006).
\bibitem{Ang2012} Y.S. Ang, Z. Ma, and C. Zhang, Scientific Reports \textbf{2}, 1013 (2012).
\bibitem{Hsieh2008} D. Hsieh, D. Qian, L. Wray, Y. Xia, Y.S. Hor, R.J. Cava, and M.Z. Hasan, Nature \textbf{452}, 970 (2008). 
\bibitem{Zhang2009N} H. Zhang, C.X. Liu, X.L. Qi, X. Dai, Z. Fang, and S.C. Zhang, Nature Phys. \textbf{5}, 438 (2009). 
\bibitem{Chen2009} Y.L. Chen, J.G. Analytis, J.H. Chu, Z.K. Liu, S.K. Mo, X.L. Qi, H.J. Zhang, D.H. Lu, X. Dai, Z. Fang, S.C. Zhang, I.R. Fisher, Z. Hussain, and Z.X. Shen, Science \textbf{325}, 178 (2009). 
\bibitem{Hsieh2009N} D. Hsieh, Y. Xia, D. Qian, L. Wray, J.H. Dil, F. Meier, J. Osterwalder, L. Patthey, J.G. Checkelsky, N.P. Ong, A.V. Fedorov, H. Lin, A. Bansil, D. Grauer, Y.S. Hor, R.J. Cava, and M.Z. Hasan, Nature \textbf{460}, 1101 (2008). 
\bibitem{Heck2011} B. van Heck, F. Hassler, A.R. Akhmerov, and C.W.J. Beenakker, Phys. Rev. B \textbf{84}, 180502 (2011).
\bibitem{Fu2009} L. Fu and C.L. Kane, Phys. Rev. B \textbf{79}, 161408(R) (2009).
\bibitem{Rademaker2011} L. Rademaker, J. Zaanen, and H. Hilgenkamp, Phys. Rev. B \textbf{83}, 012504 (2011).
\bibitem{Veldhorst2012PRB} M. Veldhorst, C.G. Molenaar, C.J.M. Verwijs, H. Hilgenkamp, and A. Brinkman, Phys. Rev. B \textbf{86}, 024509 (2012).
\bibitem{Veldhorst2013} M. Veldhorst, \textit{et al.}, Phys. Stat. Sol. RRL , 2013.
\bibitem{Spielman2000} I.B. Spielman, J.P. Eisenstein, L.N. Pfeiffer, and K.W. West, Phys. Rev. Lett. \textbf{84}, 5808 (2000).
\bibitem{Kulik1970} I.O. Kulik, JETP \textbf{30}, 944 (1970).
\bibitem{Beenakker2012} C.W.J. Beenakker, Annu. Rev. Condens. Matter Phys. \textbf{4} 113-36 (2013).
\bibitem{Hasan2010} M.Z. Hasan and C.L. Kane, Rev. Mod. Phys, \textbf{82}, 3045 (2010). 
\bibitem{Sato2010} K. Sato, D. Loss, Y. Tserkovnyak, Phys. Rev. Lett. \textbf{105}, 226401 (2010).
\bibitem{Konig2007} M. K\"onig, S. Wiedmann, C. Br\"une, A. Roth, H. Buhmann, L.W. Molenkamp, X.L. Qi, and S.C. Zhang, Science \textbf{318}, 766 (2007).
\bibitem{Brune2011} C. Br\"une, C.X. Liu, E.G. Novik, E.M. Hankiewicz, H. Buhmann, Y.L. Chen, X.L. Qi, Z.X. Shen, S.C. Zhang, and L.W. Molenkamp, Phys. Rev. Lett. \textbf{106}, 126803 (2011).
\end{thebibliography}
\end{document}